\documentclass[twocolumn,aps,showpacs,preprintnumbers,amsmath,amssymb, superscriptaddress]{revtex4-1}
\pdfoutput=1

\usepackage{bm}
\usepackage{graphicx}
\usepackage{color}
\usepackage{ulem}

\renewcommand{\vec}{\mathbf}

\begin{document}

\title{Observation of non-Fermi liquid behavior in hole-doped LiFe$_{1-x}$V$_x$As}

\author{L. Y. Xing}\thanks{These authors contributed equally to this work}
\affiliation{Beijing National Laboratory for Condensed Matter Physics, and Institute of Physics, Chinese Academy of Sciences, Beijing 100190, China}
\author{X. Shi}\thanks{These authors contributed equally to this work}
\affiliation{Beijing National Laboratory for Condensed Matter Physics, and Institute of Physics, Chinese Academy of Sciences, Beijing 100190, China}
\author{P. Richard}\email{p.richard@iphy.ac.cn}
\affiliation{Beijing National Laboratory for Condensed Matter Physics, and Institute of Physics, Chinese Academy of Sciences, Beijing 100190, China}
\affiliation{Collaborative Innovation Center of Quantum Matter, Beijing, China}
\author{X. C. Wang}\email{wangxiancheng@iphy.ac.cn}
\affiliation{Beijing National Laboratory for Condensed Matter Physics, and Institute of Physics, Chinese Academy of Sciences, Beijing 100190, China}
\author{Q. Q. Liu}
\affiliation{Beijing National Laboratory for Condensed Matter Physics, and Institute of Physics, Chinese Academy of Sciences, Beijing 100190, China}
\author{B. Q. Lv}
\affiliation{Beijing National Laboratory for Condensed Matter Physics, and Institute of Physics, Chinese Academy of Sciences, Beijing 100190, China}
\author{J.-Z. Ma}
\affiliation{Beijing National Laboratory for Condensed Matter Physics, and Institute of Physics, Chinese Academy of Sciences, Beijing 100190, China}
\author{B. B. Fu}
\affiliation{Beijing National Laboratory for Condensed Matter Physics, and Institute of Physics, Chinese Academy of Sciences, Beijing 100190, China}
\author{L.-Y. Kong}
\affiliation{Beijing National Laboratory for Condensed Matter Physics, and Institute of Physics, Chinese Academy of Sciences, Beijing 100190, China}
\author{H. Miao}
\affiliation{Condensed Matter Physics and Materials Science Department, Brookhaven National Laboratory, Upton, New York 11973, USA}
\author{T. Qian}
\affiliation{Beijing National Laboratory for Condensed Matter Physics, and Institute of Physics, Chinese Academy of Sciences, Beijing 100190, China}
\author{T. K. Kim}
\affiliation{Diamond Light Source, Harwell Campus, Didcot, OX11 0DE, United Kingdom}
\author{M. Hoesch}
\affiliation{Diamond Light Source, Harwell Campus, Didcot, OX11 0DE, United Kingdom}
\author{H. Ding}\email{dingh@iphy.ac.cn}
\affiliation{Beijing National Laboratory for Condensed Matter Physics, and Institute of Physics, Chinese Academy of Sciences, Beijing 100190, China}
\affiliation{Collaborative Innovation Center of Quantum Matter, Beijing, China}
\author{C. Q. Jin}\email{cqjin@iphy.ac.cn}
\affiliation{Beijing National Laboratory for Condensed Matter Physics, and Institute of Physics, Chinese Academy of Sciences, Beijing 100190, China}
\affiliation{Collaborative Innovation Center of Quantum Matter, Beijing, China}

\date{\today}

\begin{abstract}
We synthesized a series of V-doped LiFe$_{1-x}$V$_x$As single crystals. The superconducting transition temperature $T_c$ of LiFeAs decreases rapidly at a rate of 7 K per 1\% V. The Hall coefficient of LiFeAs switches from negative to positive with 4.2\% V doping, showing that V doping introduces hole carriers. This observation is further confirmed by the evaluation of the Fermi surface volume measured by angle-resolved photoemission spectroscopy (ARPES), from which a 0.3 hole doping per V atom introduced is deduced. Interestingly, the introduction of holes does not follow a rigid band shift. We also show that the temperature evolution of the electrical resistivity as a function of doping is consistent with a crossover from a Fermi liquid to a non-Fermi liquid. Our ARPES data indicate that the non-Fermi liquid behavior is mostly enhanced when one of the hole $d_{xz}/d_{yz}$ Fermi surfaces is well nested by the antiferromagnetic wave vector to the inner electron Fermi surface pocket with the $d_{xy}$ orbital character. The magnetic susceptibility of LiFe$_{1-x}$V$_x$As suggests the presence of strong magnetic impurities following V doping, thus providing a natural explanation to the rapid suppression of superconductivity upon V doping.
\end{abstract}



\maketitle

\section{Introduction}

The Fe-based superconductors are multi-band systems with a Fermi surface (FS) composed of several pockets with different orbital characters \cite{RichardRoPP2011}. The complexity of the FS allows different low-energy scattering processes that can stabilize long-range magnetic order and that have even been proposed to induce superconducting (SC) pairing in these compounds \cite{MazinPRL2008,Ding_EPL,Graser_NJP2009}. A recent report on Co-doped LiFeAs reveals a direct correlation between FS nesting and manifestations of non-Fermi liquid (NFL) behavior in the absence of a clear quantum critical point \cite{Dai_PRX5}, \textit{i. e.} a phase transition at zero temperature. Indeed, the largest deviation of the temperature exponent in the power law used to fit the temperature dependence of the electrical resistivity is found for a doping corresponding to a good nesting of the largest hole FS pocket with $d_{xy}$ character with the electron pockets. Unfortunately, whether the relevant scattering causing the NFL behavior is orbital-dependent is unclear. A natural way to settle this issue is to hole-dope LiFeAs and search for similar scattering but with FS pockets carrying different orbital characters. Although attempts to hole-dope the BaFe$_2$As$_2$ system by substituting Fe by lighter 3$d$ elements have been done using Mn \cite{Y_Liu_PhysicaC470, Kim_PRB84, Thaler_PRB84,Pandey_PRB84} and Cr \cite{Sefat_PRB79,Budko_PRB80,Colombier_SST23,Marty_PRB83,Clancy_PRB85}, literature lacks of similar study on LiFeAs, and in either cases V has not been used to induce hole doping.  

In this paper we report the synthesis and characterization of V-doped LiFeAs. The SC transition temperature $T_c$ of LiFeAs is dramatically suppressed by V doping at a rate of 7~K per 1\% V, which is much faster than for Co/Ni/Cu doping. Our Hall coefficient $R_H$ and angle-resolved photoemission spectroscopy (ARPES) measurements indicate that each V dopant releases 0.3~hole carrier. With the V doping level increasing, we observe a crossover from FL to Non-FL to FL behavior, which we explain by increased inter-orbital scattering mediated by low-energy spin fluctuations when a good nesting is found between the inner $d_{xy}$ electron FS pocket at the M point and the outer $d_{xz}/d_{yz}$ hole FS pocket at the $\Gamma$ point. Also, our susceptibility measurements suggest that V doping gives rise to magnetic impurities, thus providing a simple explanation for the rapid suppression of $T_c$ upon V doping.

\section{Experiment}

The single crystals of LiFe$_{1-x}$V$_x$As used in this study have been synthesized using the self-flux method and the details can be found elsewhere \cite{XingJPCM26}. Li$_3$As and Fe$_{1-x}$V$_x$As were first synthesized as precursors by a solid state reaction method. To synthesize Li$_3$As, Li lump and As powder were sealed into a Ti tube under 1 atm Ar and sintered at 650 $^{\circ}$C for 10 hours. Fe$_{1-x}$V$_x$As was obtained by mixing Fe, V and As powders, pressing into pieces and heating at 700 $^{\circ}$C for 15 hours in an evacuated quartz tube. We then mixed Li$_3$As, Fe$_{1-x}$V$_x$As and As together with the ratio of Li$_3$As:Fe$_{1-x}$V$_x$As:As=1:0.9:1.1, and put into an alumina crucible, sealed into a Nb tube under 1 atm Ar atmosphere and sealed into a quartz tube. The mixture were first heated to 1100 $^{\circ}$C for 50 hours and then slowly cooled down to 750 $^{\circ}$C at the rate of 2 $^{\circ}$C/h to grow the single crystals. The chemical compositions of the LiFe$_{1-x}$V$_x$As single crystals used hereafter were determined by energy dispersive X-ray spectroscopy (EDX) with an experimental accuracy of about 10\%.

The samples were characterized by X-ray powder diffraction (XRD) in the 10$^o$ to 80$^o$ range using a Philips XÕpert  diffractometer from and a scanning rate of 10$^o$ per minute. The dc magnetic susceptibility was measured by using a vibrating sample magnetometer (VSM) with magnetic field of 30 Oe for characterizing the superconducting transition, and of 1 T for study the effective magnetic moment. Resistivity and Hall coefficient measurements were performed using a six-probe technique on a Quantum Design instrument physical property measurement system (PPMS) with a magnetic field up to 6 T. We performed ARPES measurements at the Dreamline beamline of Shanghai Synchrotron Radiation Facility and at the I05 beamline of Diamond Light Source using a Scienta D80 and a Scienta R4000 analyzers, respectively. The energy and angular resolutions have been set to 10 meV and 0.2$^{\circ}$, respectively. All measurements have been recorded at 20 K with 51 eV photons ($k_z\approx 0$ \cite{Miao_NCOMM6}) in $s$ and $p$ polarization configurations.

\begin{figure}[!t]
\begin{center}
\includegraphics[width=\columnwidth]{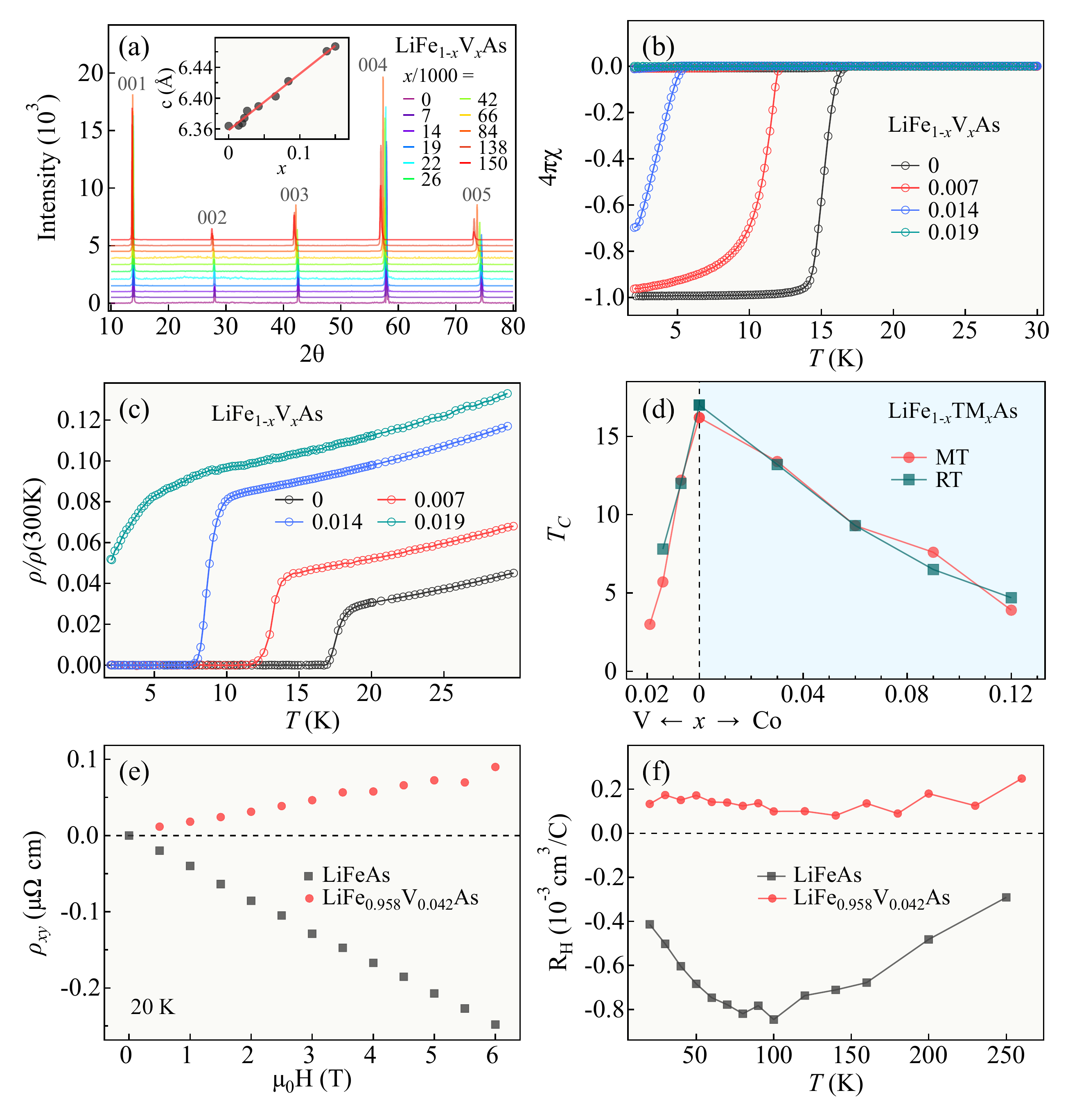}
\end{center}
\caption{\label{Hall}(Color online) (a) XRD patterns of LiFe$_{1-x}$V$_x$As single crystals. The inset shows the dependence of the lattice constant $c$ versus the doping level $x$, fit to a linear line (red). (b) and (c) Temperature dependence of the magnetic susceptibility and the normalized resistivity $\rho/\rho(\textrm{300 K})$ of LiFe$_{1-x}$V$_x$As single crystals. (d) Evolution of $T_c$ as a function of V and Co doping in LiFe$_{1-x}$TM$_x$As (TM = V and Co) single crystals. MT and RT represent $T_c$ determined from the magnetic susceptibility and from the resistivity, respectively. The data for LiFe$_{1-x}$Co$_x$As are extracted from Ref. \cite{XingJPCM26}, copyright \copyright~(2014) by IOP Publishing. (e) Magnetic field dependence of the Hall resistivity $\rho_{xy}$ of LiFeAs and LiFe$_{0.958}$V$_{0.042}$As single crystals at 20 K. (f) Temperature dependence of the Hall coefficient $R_H$ of LiFeAs and LiFe$_{0.958}$V$_{0.042}$As single crystals. }
\end{figure}

\section{Results and discussion}

Fig. \ref{Hall}(a) shows the XRD patterns of LiFe$_{1-x}$V$_x$As single crystals. Only $00l$ peaks are observed and the full-width-at-half-maxima are smaller than $0.2^{\circ}$, suggesting high single crystal quality. The inset shows the doping evolution of the $c$ axis, which is consistent with Vegard's law. While superconductivity in most Fe-based superconductors occurs upon carrier doping, a $T_c$ of 18 K is observed in pristine LiFeAs \cite{Wang_SSC148,Tapp_PRB78}. The $T_c$ of LiFeAs decreases linearly upon electron doping with Co, Ni or Cu \cite{Pitcher_JACS132,XingJPCM26}. Figs. \ref{Hall}(b) and \ref{Hall}(c) show the temperature dependence of the magnetic susceptibility and the resistivity for LiFe$_{1-x}$V$_x$As with $x<0.019$. All LiFe$_{1-x}$V$_x$As samples with $x < 0.019$ exhibit diamagnetic susceptibility and SC zero resistivity. The SC transition $T_c$ decreases from 16~K to 6~K with $x$ increasing to 0.014. Superconductivity disappears at $x = 0.019$, as shown in Fig. \ref{Hall}(c). As displayed in Fig. \ref{Hall}(d), $T_c$ decreases almost linearly with V doping at a rate of 7~K per 1\%V, which is much faster than for electron doping with Co/Ni/Cu \cite{Canfield_PRB80,AF_Wang_PRB88}. Indeed, 1\% doping with Co and Ni in the LiFe$_{1-x}$(Co/Ni)$_x$As system suppresses $T_c$ by 1 K and 2.2 K, respectively. Cu doping suppresses $T_c$ at a rate similar as Ni doping but the FS almost does not change. It was proposed that the suppression of $T_c$ by Cu dopant is mainly due to its strong impurity scattering effect \cite{XingJPCM26}.

We further study the effect of V doping in the LiFeAs system from Hall resistivity measurements. Fig.~\ref{Hall}(e) compares the Hall resistivity $\rho_{xy}$ of LiFeAs and LiFe$_{0.958}$V$_{0.042}$As versus applied magnetic field measured at 20 K. In contrast to LiFeAs, for which the slope of $d\rho_{xy}/dH$ is negative, the slope of $d\rho_{xy}/dH$ is positive for LiFe$_{0.958}$V$_{0.042}$As, suggesting that the substitution of Fe by V dopes the system with holes. The temperature dependence of the Hall coefficient $R_H$ for these samples is shown in Fig. \ref{Hall}(f). For LiFeAs, $R_H$ is negative with a minimum value at about 100 K, which is consistent with a previous work \cite{Heyer_PRB84}. However, for the LiFe$_{0.958}$V$_{0.042}$As sample, $R_H$ is positive with a smaller temperature dependence within the measured temperature range. This sign reversal of $R_H$ is an explicit indication of hole doping with the introduction of V.

\begin{figure}[!t]
\begin{center}
\includegraphics[width=\columnwidth]{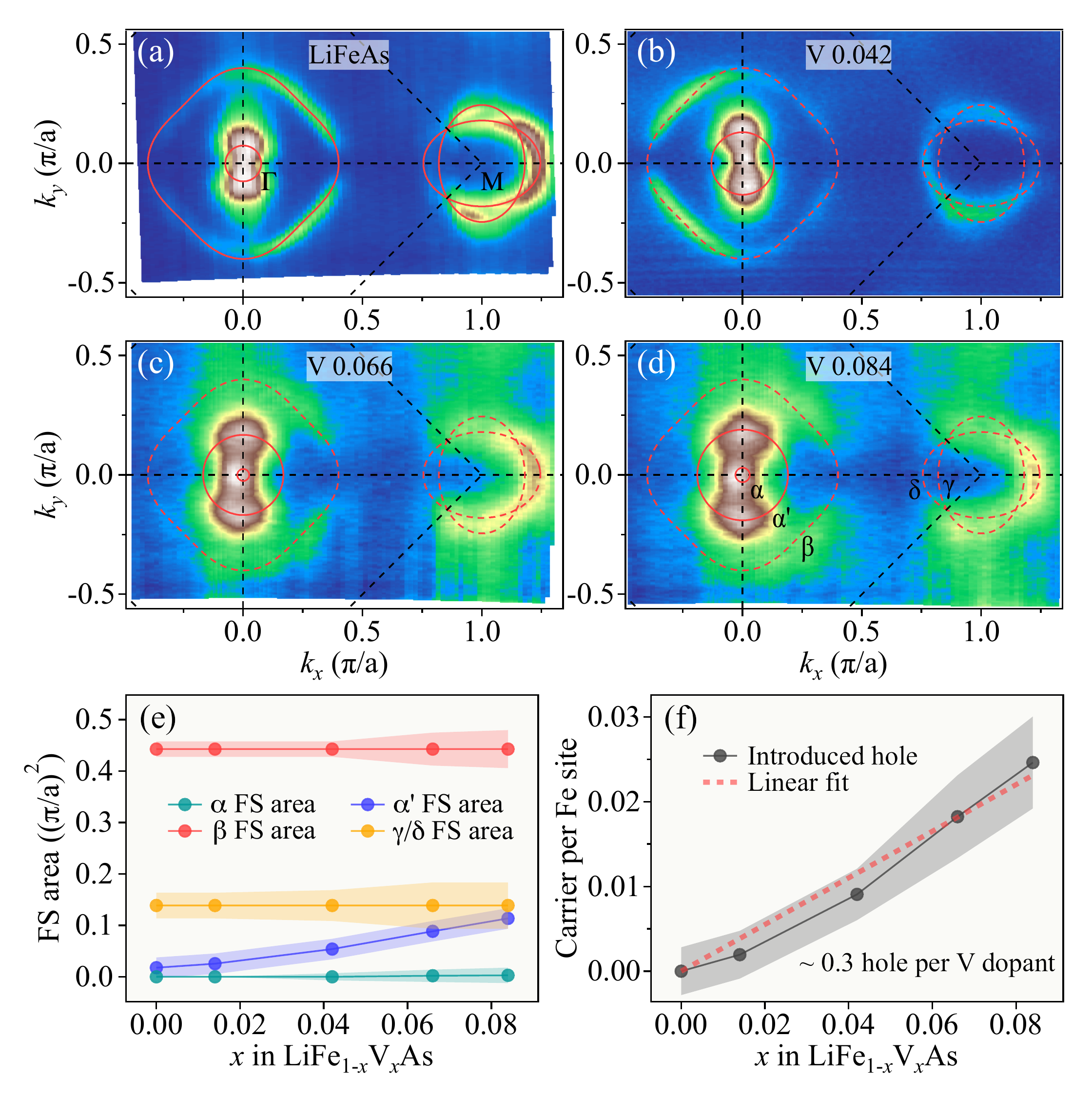}
\end{center}
\caption{\label{FS}(Color online) (a)-(d) ARPES FS intensity maps of LiFe$_{1-x}$V$_x$As for $x$ = 0, 0.042, 0.066 and 0.084, respectively. The maps were recorded with $s$ polarized 51 eV photons and integrated within a 10 meV energy window centered at $E_F$. The solid red lines appended on the plots are extracted FS contours at each doping. The dashed red lines correspond to data extracted in LiFeAs. (e) Calculated area of different FS pockets as a function of doping. (f) Number of carriers introduced by the Fe~$\rightarrow$~V substitution deduced from the FS volume. The dashed pink line is a linear fit showing that 0.3 hole is added to the system for each V dopant.} 
\end{figure}

\begin{figure}[!t]
\begin{center}
\includegraphics[width=\columnwidth]{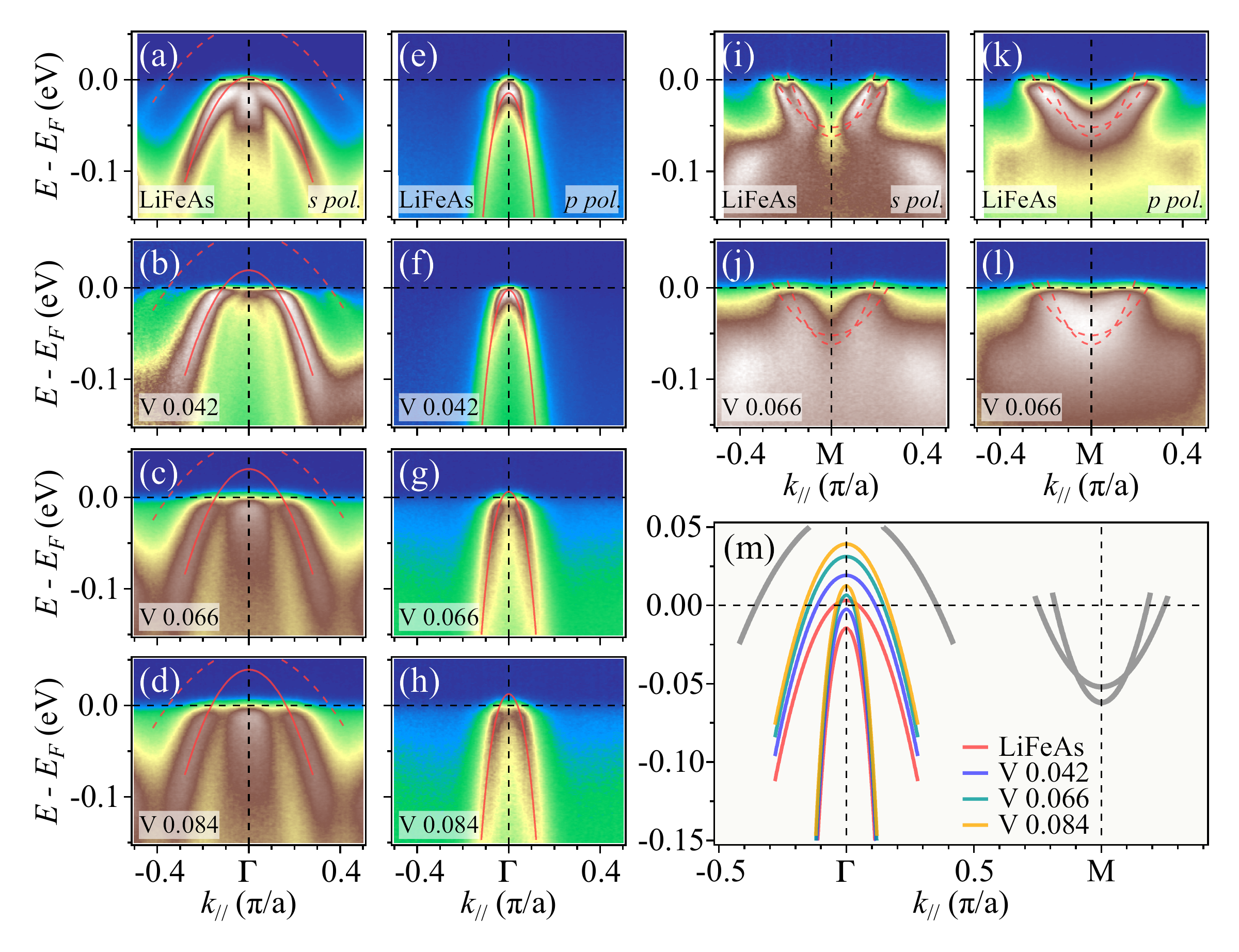}
\end{center}
\caption{\label{band}(Color online) (a)-(d) ARPES intensity plots near $\Gamma$ of LiFe$_{1-x}$V$_x$As recorded with $s$ polarized light for $x$ = 0, 0.042, 0.066 and 0.084, respectively. (e)-(h) Same as (a)-(d) but recorded with $p$ polarized light. (i)-(j) Intensity plots near M for $x$ = 0 and 0.066 samples, obtained at 20 K along $\Gamma$-M direction, with $s$ polarized 51 eV photons. (k)-(l) Same as (i)-(j) but using $p$ polarized photons. (m) Comparison of the band dispersions. The outer hole-like band and the electron-like bands, in grey, do not change within the experimental uncertainties.}
\end{figure}

\begin{figure*}[!t]
\begin{center}
\includegraphics[width=\textwidth]{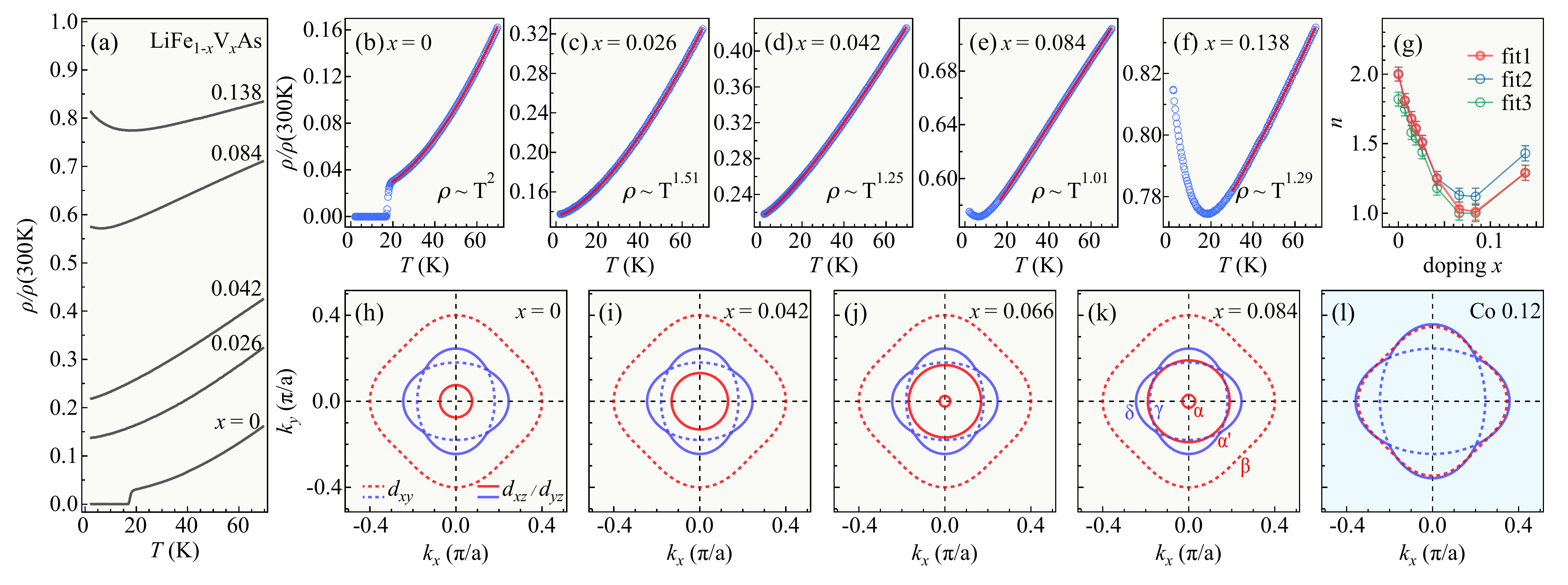}
\end{center}
\caption{\label{resistivity}(Color online) (a) Temperature dependence of the normalized resistivity $\rho/\rho(\textrm{300 K})$ for five selected LiFe$_{1-x}$V$_x$As single crystals. (b)-(f) Same as (a) both with a zoom on each resistivity curve. For each sample, $\rho/\rho(\textrm{300 K})$ is fitted using $\rho=\rho_0+AT^n$.  (g) Exponent $n$ as a function of $x$. The red circles (fit1) represent $n$ values obtained from the fitted curves shown in Figs. 3(b)-3(f). The cyan circles (fit2) are the values obtained in the temperature range that starts by the metal-insulator transition or $T_c$. The green circles (fit3) are the values obtained in the 30-70 K temperature range. (h)-(k) Evolution of the hole FSs (red) and electron FSs (blue) as a function of $x$ in LiFe$_{1-x}$V$_x$As. The electron FSs pockets have been shifted to the $\Gamma$ point for a better comparison. Except for the electron-like pockets and the large $\beta$ hole-like pocket are treated as doping-independent, in agreement with the experimental results. (l)~Same as (h)-(k) but for LiFe$_{0.88}$Co$_{0.12}$As (extracted from Ref. \cite{Dai_PRX5}), which shows the most pronounced NFL behavior in the LiFe$_{1-x}$Co$_x$As series.}
\end{figure*}

In order to confirm and quantify the introduction of hole carriers in the V-doped LiFeAs samples, we recorded ARPES data at selected sample dopings. We show in Figs. \ref{FS}(a)-\ref{FS}(d) the FS measured for the V doping levels $x = 0$, 0.042, 0.066 and 0.084, respectively. For each sample, hole FS pockets at $\Gamma$ and electron FS pockets at M are observed, as indicated by the red contours extracted from the ARPES intensity maps. The FS topology is typical of that of most iron-pnictides \cite{RichardRoPP2011}. We note that the band structure of highly V-doped samples ($x$ = 0.066, 0.084), and in particular the $\beta$ FS sheet ($d_{xy}$ character), is not as sharp as that of pristine LiFeAs. This is intrinsically due to strong impurity scattering. The FS can be well approximated nevertheless, especially from the trend observed in the doping evolution at low-doping, where the bands are clearly identifiable. Surprisingly, we find that while the $\alpha$ (not crossing the Fermi level ($E_F$) in LiFeAs) and $\alpha'$ FS pockets get enlarged with the V concentration increasing, the other FSs barely change with doping. Consequently, we use the FSs extracted in LiFeAs to approximate the $\beta$ FS pocket and the electron-like pockets. Obviously, these FS contours are consistent with the FS intensity patterns at all dopings, thus validating our approximation. As shown in Figs. \ref{FS}(e) and \ref{FS}(f), error bars are used to account for our limited accuracy in determining the FSs precisely at high doping.

According to the Luttinger theorem, the electronic carrier concentration is determined by the algebraic sum of all the FS areas. The calculated area of each FS pocket as a function of doping is shown in Fig. \ref{FS}(e), and the deduced hole carrier concentration is illustrated in Fig. \ref{FS}(f). Based on a previous ARPES report \cite{Brouet_PRB93} exhibiting only small warping of the electronic structure of LiFeAs along the $k_z$ direction, the carrier concentration was evaluated by considering the data recorded in the $k_z=0$ plane only. As expected from the Hall coefficient data, the substitution of Fe by V leads to hole doping. However, we conclude from a simple linear fit of the introduced free carrier concentration as a function of the V content that only 0.3 hole carrier is introduced per substituting V dopant. We note that neglecting the slight three-dimensionality of the FS leads to an uncertainty smaller than 0.05 holes. The experimental doping of $0.3 \pm 0.1$ is ten times smaller than the ``nominal" value, which contrasts with the Co and Ni-doped cases \cite{Dai_PRX5,Pitcher_JACS132}. Co doping into LiFeAs can be approximated by a rigid band shift with one free electron introduced for each Co atom. The electron FS pockets become larger while the hole FS pockets shrink upon Co doping in LiFeAs, with their algebraic sum respecting the Luttinger theorem \cite{Dai_PRX5}. The situation is quite different in V-doped LiFeAs. As indicated in Fig. \ref{FS}(e), the change in the free carrier concentration is practically related only to the $\alpha'$ FS pocket, thus indicating that V doping does not follow a rigid band shift.

This unusual band structure evolution with V doping is further illustrated in Fig. \ref{band}. We use $s$ polarized and $p$ polarized incident light to highlight the bands carrying different orbital characters. As shown in Figs. \ref{band}(a)-\ref{band}(h), the inner hole bands at $\Gamma$ ($d_{xz}/d_{yz}$) shift towards lower binding energy upon V doping, as expected. In contrast, the dispersions of the outer hole band at $\Gamma$ [$d_{xy}$, see Figs. \ref{band}(a)-\ref{band}(d)] and of the electron bands at M [Figs. \ref{band}(i)-\ref{band}(l)] are practically unchanged, at least within the current experimental uncertainties. This is summarized in Fig. \ref{band}(m). The reason for this dichotomy of band filling behavior in V-doped LiFeAs is unclear and goes beyond the scope of this paper.

\begin{figure}[!t]
\begin{center}
\includegraphics[width=\columnwidth]{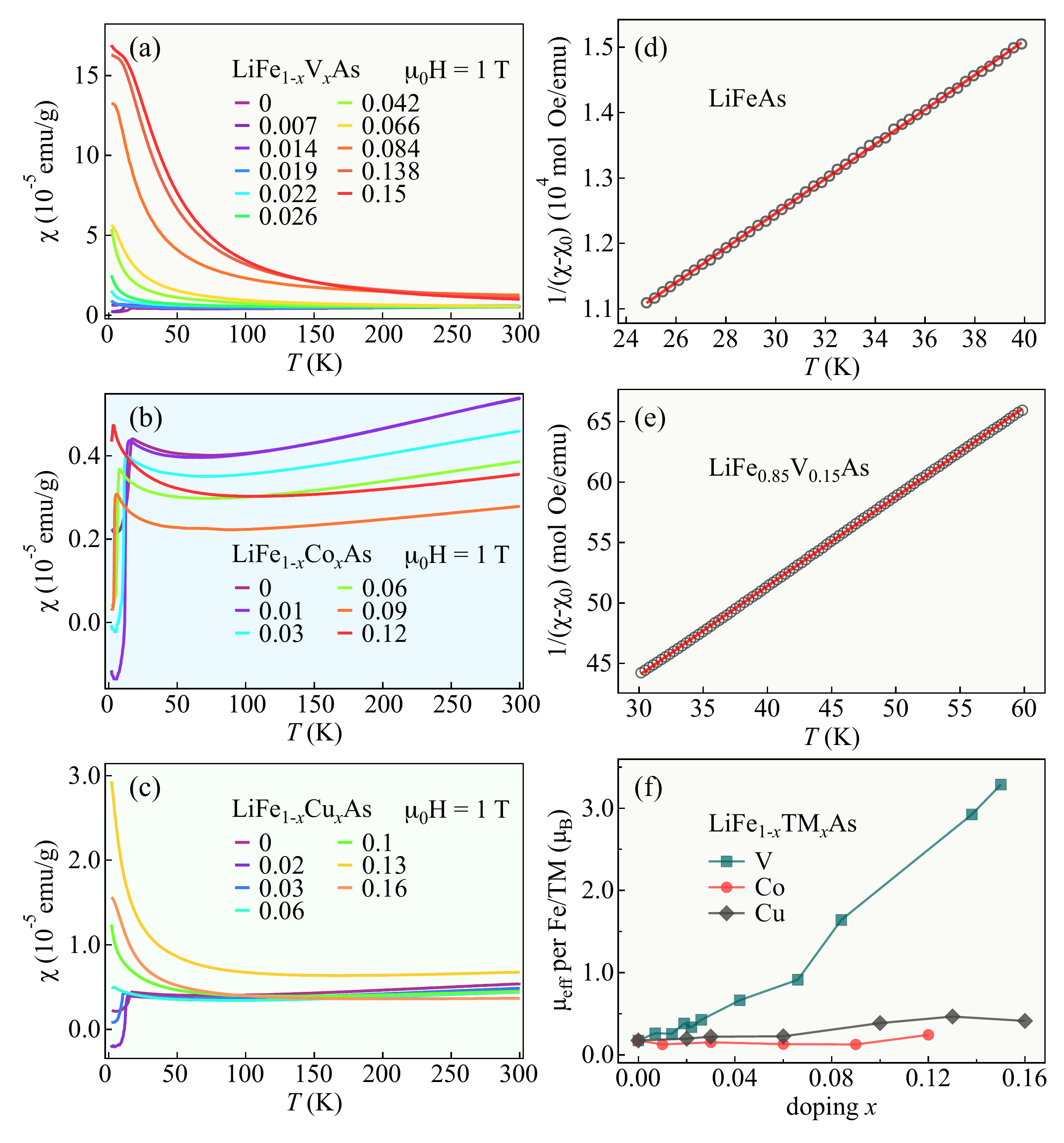}
\end{center}
\caption{\label{susceptibility}(Color online) Temperature dependence of the magnetic susceptibility in a 1 T magnetic field of (a) LiFe$_{1-x}$V$_x$As, (b) LiFe$_{1-x}$Co$_x$As and (c) LiFe$_{1-x}$Cu$_x$As single crystals. (d) and (e) Magnetic susceptibility fitted according to the Curie-Weiss law $1/(\chi-\chi_0)=(T-\theta)/C$. (f) Effective moment per Fe/TM site as a function of the doping concentration $x$ for LiFe$_{1-x}$TM$_x$As (TM = V, Co and Cu) single crystals.}
\end{figure}

A recent report on Co-doped LiFeAs indicates that the improvement upon doping of the nesting conditions between the hole and electron FS pockets leads to a Fermi liquid (FL) to a NFL behavior \cite{Dai_PRX5}. As the size of the $\alpha'$ FS pocket increases with V doping, whereas the electron FS pockets remain nearly unchanged, one can wonder if FS nesting can also lead to a NFL in this system. To check this scenario, we measured the resistivity of the LiFe$_{1-x}$V$_x$As samples and fitted the data with $\rho=\rho_0+AT^n$, as shown in Fig. \ref{resistivity}. To minimize contamination from the semiconducting transition at low temperature for highly-doped samples, we compare the results obtained with three different temperature ranges. In the first series of fits [fit1 in Fig. \ref{resistivity}(g)], the lowest temperature in chosen as lower as possible, but with the minimum of the resistivity curve outside of the fitting range. The corresponding temperature range fits are given by the red curves in Figs. \ref{resistivity}(b)-\ref{resistivity}(f). In the second series [fit2 in Fig. \ref{resistivity}(g)], the fit starts at the semiconducting transition. Finally, for the last series of fit [fit3 in Fig. \ref{resistivity}(g)], the resistivity was fitted in the 30 - 70 K temperature range for all dopings. Despite small discrepancies, the doping evolution of the exponent $n$, displayed in Fig. \ref{resistivity}(g), has the same trends for all the fitting methods. 

As with Co-doped LiFeAs \cite{Dai_PRX5}, the exponent $n$ deviates from 2 with the introduction of V dopants, suggesting a NFL regime. We emphasize that this is already true at low doping, away from the semiconducting transition. Interestingly, $n$ reaches a minimum of 1.01 for a doping of $x = 0.084$, before increasing again towards 2 with further doping. We caution that although the precise values of the exponent $n$ are questionable at high doping, the trend is clear. As with Co-doped LiFeAs, this behavior strongly suggests that V-doped LiFeAs is host to low-energy fluctuations enhanced with the nesting conditions. The observation of a NFL behavior without quantum critical point in both Co-doped and V-doped LiFeAs is consistent with a recent theoretical study emphasizing on the impact of nesting on the temperature evolution of the electrical conductivity \cite{Setty_PRB93}. We caution though that the linear resistivity in the cuprates was previously derived in the context of a FL with a large density-of-states (like a van Hove singularity) near $E_F$ \cite{Kastrinakis_PhysicaC340,Kastrinakis_PRB71}, a condition that also applies to Fe-based superconductors near good nesting conditions.

As shown in Fig. \ref{resistivity}(k), a good nesting is found at $x = 0.084$ between the $\alpha'$ FS pocket, which carries a $d_{xz}/d_{yz}$ orbital character, and the inner electron FS pocket ($\gamma$), mainly derived from the $d_{xy}$ orbital. This situation is symmetrical to the one found in Co-doped LiFeAs, for which the lowest $n$ coincides with the best nesting conditions between the outer hole FS pocket ($\beta$) with a $d_{xy}$ orbital character and the outer electron FS pocket ($\delta$), which carries a dominant $d_{xz}/d_{yz}$ orbital character [Fig. \ref{resistivity}(l)]. It is thus fair to say that the NFL behavior is mainly due to low-energy inter-orbital scattering with the scattering vector $\vec{Q}$ corresponding to the wavevector between $\Gamma$ and M. Such experimental observation is consistent with theoretical calculations indicating that the coupling to the wave vector $\vec{Q}$ is favored between different orbital states \cite{J_Zhang_PRB79}. We caution that unlike the assumption made in Ref. \cite{J_Zhang_PRB79}, this scattering is not obviously correlated to the SC pairing.

However, our data suggest that the magnetic properties of LiFe$_{1-x}$V$_x$As go beyond simple FS nesting. When $x>0.066$, LiFe$_{1-x}$V$_x$As displays a semiconducting behavior at low temperature similar to the one observed in Cu-doped Fe-based superconductors. A weak semiconducting behavior emerges at $x\sim 0.13$ for LiFe$_{1-x}$Cu$_x$As \cite{XingIJMPB29} and at $x\sim 0.033$ for NaFe$_{1-x}$Cu$_x$As \cite{AF_Wang_PRB88}. A crossover from metallic to semiconducting behavior with Cu doping has also been observed in BaFe$_{1-x}$Cu$_x$As \cite{Ni_PRB82}, SrFe$_{1-x}$Cu$_x$As \cite{Yan_PRB87} and Fe$_{1.01-x}$Cu$_x$Se \cite{WilliamsJPCM21}, and it was proposed that Cu dopants in these materials behave like strong scatterers inducing a metal-insulator transition due to Anderson localization \cite{Chadov_PRB81}. Since the critical doping for the metal-semiconductor crossover is even lower in LiFe$_{1-x}$V$_x$As, \textit{i.e.} $x=0.066$ in LiFe$_{1-x}$V$_x$As as compared with $x=0.4$ in LiFe$_{1-x}$Co$_x$As and $x=0.13$ in LiFe$_{1-x}$Cu$_x$As, we also assume that the V impurities are strong scattering centers.

We also measured the magnetic susceptibility under magnetic field up to 1 T. The amplitude of the magnetic susceptibility in LiFe$_{1-x}$V$_x$As [Fig. \ref{susceptibility}(a)] is about one order of magnitude higher than for LiFe$_{1-x}$Co$_x$As [Fig. \ref{susceptibility}(b)] and LiFe$_{1-x}$Cu$_x$As [Fig. \ref{susceptibility}(c)]. As shown in Figs. \ref{susceptibility}(d) and \ref{susceptibility}(e), respectively, the susceptibilities of LiFeAs and LiFe$_{0.85}$V$_{0.15}$As follow the Curie-Weiss~law $1/(\chi-\chi_0)=(T-\theta)/C$ at low temperature, where $\theta$ is the~Curie~temperature. The Curie-Weiss law allows us to extract an effective magnetic moment $\mu_{eff}$. The results as a function of the content $x$, displayed in Fig.~\ref{susceptibility}(f), indicate that while $\mu_{eff}$ varies only slightly upon Co and Cu doping, it is greatly enhanced with V doping.

The enhancement of $\mu_{eff}$ by Mn dopants has been reported in the NaFeAs system \cite{Deng_NJP16}, where 4\%-Mn dopants increase the effective moment of Na(Fe$_{0.97-x}$Co$_{0.03}$Mn$_x$)As from about 0.4 to 1.8 $\mu_B$ per Fe/Co/Mn site. The Mn dopants in Na(Fe$_{0.97-x}$Co$_{0.03}$Mn$_x$)As are considered to be magnetic impurities and spectroscopic signature of Kondo screening on the Mn adatoms has been observed by scanning tunneling spectroscopy \cite{Deng_NJP16, PhysRevB.89.214515}. Accordingly, although our experiment does not allow us to conclude unambiguously that the magnetic moments develop on the V sites, we speculate that the V dopants in our study are magnetic impurities because of the enhancement of $\mu_{eff}$ with V substitution. 

The effective magnetic moment of LiFeAs is enhanced to about 3.3 $\mu_B$ by 16\% V dopants, which is unexpectedly higher than for usual Fe-based superconductors. Yet, an even higher effective magnetic moment of about 4 $\mu_B$ has been reported in Fe$_{1+x}$Te and explained by a Kondo-type behavior (\textit{i.e.} with local magnetic moments entangled with itinerant electrons) \cite{PhysRevLett.107.216403, PhysRevB.80.214514, PhysRevB.93.024504}. Similarly, V dopants are likely to give rise to magnetic impurities and induce entanglement between the local magnetic moments and the itinerant electrons. Here the local magnetic moments may come from either magnetic impurities or from Fe, but further discussion is beyond the scope of this paper. Finally, we note that the observation of strong magnetic moments offers one possible explanation to the rapid suppression of $T_c$ in LiFe$_{1-x}$V$_x$As.

\section{Summary}

In summary, we synthesized and characterized single crystals of LiFe$_{1-x}$V$_x$As. Using Hall coefficient and ARPES measurements, we showed that the substitution of Fe by V dopes the system with holes at a rate of 0.3~hole/V. We showed from the temperature dependence of the resistivity data that the system exhibits a NFL behavior that is enhanced around doping where our ARPES measurements indicate a good FS nesting between FS pockets carrying different orbital characters, suggesting that the NFL behavior is induced by low-energy inter-orbital scattering by the antiferromagnetic wave vector. Finally, the magnetic susceptibility of LiFe$_{1-x}$V$_x$As suggest that a strong magnetic moments develop following V doping, thus providing a natural explanation for the rapid suppression of superconductivity found upon doping. 

\section*{Acknowledgement}

We acknowledge Jiang-Ping Hu for useful discussions. This work was supported by grants from MOST (2011CBA001000, 2011CBA00102, 2012CB821403, 2013CB921700 and 2015CB921301) and NSFC (11004232, 11034011/A0402, 11234014, 11274362, 11474330 and 11474344) of China. We acknowledge Diamond Light Source for time on beamline I05 under proposal SI11452.

\bibliography{biblio_V_LiFeAs}

\end{document}